%% file: main.tex
\documentclass[conference]{IEEEtran}

\usepackage{cite}
\usepackage{amsmath, amssymb, amsfonts}
\usepackage{graphicx}
\usepackage{textcomp}
\usepackage{xcolor}

\def\BibTeX{{\rm B\kern-.05em{\sc i\kern-.025em b}\kern-.08em
    T\kern-.1667em\lower.7ex\hbox{E}\kern-.125emX}}

\usepackage[linesnumbered,ruled,vlined]{algorithm2e}
\usepackage[small]{caption}
\usepackage{subfig}
\usepackage{color,soul}
\usepackage{array}
\usepackage{booktabs}
\usepackage{footnote}
\makesavenoteenv{tabular}
\makesavenoteenv{table}
\usepackage[linesnumbered,ruled,vlined]{algorithm2e}
\usepackage[noend]{algpseudocode}
\usepackage{float}
\usepackage[utf8x]{inputenc}
\usepackage{textcomp}

\usepackage{mathtools}
\DeclarePairedDelimiter\ceil{\lceil}{\rceil}

\renewcommand\baselinestretch{0.975}

\begin{document}
\title{NN-PARS: A Parallelized Neural Network Based \\  Circuit Simulation Framework}

\input{authors.tex}

\maketitle
\thispagestyle{plain}
\pagestyle{plain}

\input{abstract.tex}


\IEEEpeerreviewmaketitle

\input{introduction.tex}
\input{background.tex}

\input{method.tex}

\input{experiment.tex}

\input{conclusion.tex}

\section*{Acknowledgement}
This research was sponsored in part by a grant from the Software and Hardware Foundations (SHF) program of the National Science Foundation (NSF). 

\bibliographystyle{IEEEtran}
\bibliography{IEEEabrv,references}

\end{document}

%% file: authors.tex
\author{
\IEEEauthorblockN{Mohammad Saeed Abrishami, Hao Ge, Justin F. Calderon, Massoud Pedram, and Shahin Nazarian\\}
\IEEEauthorblockA{Ming Hsieh Department of Electrical and Computer Engineering\\
Viterbi School of Engineering, University of Southern California \\ 
Los Angeles, CA 90089 \\
\{abri442, haoge, jfcalder, pedram, shahin.nazarian\}@usc.edu}}


%% file: abstract.tex
\begin{abstract}



The shrinking of transistor geometries as well as the increasing complexity of integrated circuits, significantly aggravate nonlinear design behavior. 
This demands accurate and fast circuit simulation to meet the design quality and time-to-market constraints.  
The existing circuit simulators which utilize lookup tables and/or closed-form expressions are either slow or inaccurate in analyzing the nonlinear behavior of designs with billions of transistors. 
To address these shortcomings, we present NN-PARS, a neural network (NN) based and parallelized circuit simulation framework with optimized event-driven scheduling of simulation tasks to maximize concurrency, according to the underlying GPU parallel processing capabilities. NN-PARS replaces the required memory queries in traditional techniques with parallelized NN-based computation tasks. 
Experimental results show that compared to a state-of-the-art current-based simulation method, NN-PARS reduces the simulation time by  over two orders of magnitude in large circuits. NN-PARS also provides high accuracy levels in signal waveform calculations, with less than $2\%$ error compared to HSPICE. 
\end{abstract}

%% file: introduction.tex
\section{Introduction}\label{sec:intro}

As the CMOS transistor technologies test the limits of Moore's Law~\cite{moore-law}, 
the design flow of VLSI circuits demand increasingly more complex analysis, transformation, and verification iterations, to validate the correctness of functionality, and quality of design in terms of performance, power and signal integrity. The design flow steps need to also validate various \textit{process-voltage-temperature} (PVT) corners and operating modes such as \textit{low-power} (LP) and  \textit{high-performance} (HP) that involve increasingly nonlinear effects. 
Fast and accurate simulation is therefore crucial to help lower the number of design iterations,  speed up convergence, and consequently shorten the design turnaround time ~\cite{Kahng-ICCD2018}. 

SPICE simulators are the de facto standard tools for accurate analysis and sign-off, however they are very slow for billion-transistor circuits ~\cite{thermal-pedram-2006, dynamic-power-benini-2000}. 
Therefore, higher levels of circuit abstraction using approximation have been used to speed up simulation steps. Abstraction models are generally based on \textit{look-up-tables} (LUTs), closed-form formulations, factors or their combinations. 
The traditional voltage based models, namely \textit{nonlinear delay model} (NLDM), \textit{nonlinear power model} (NLPM), \textit{effective current source model} (ECSM~\cite{Cadence-ECSM}), and \textit{composite current source model} (CCSM~\cite{Synopsys-CCSM}) utilize LUTs for storing delay, noise or power as nonlinear functions w.r.t. physical,  structural, and environmental parameters, and depend on voltage modeling more than current modeling. Voltage based models are intuitively better choices when compared to simple closed-form formulation of nonlinear functions, however, it tends to be increasingly inaccurate in capturing signal integrity and short channel effects with the down-scaling of technologies~\cite{goel-integrity-date2008}. 
Alternatively, current based models such as \textit{Current Source Models} (CSMs) \cite{croix2003blade, Cadence-CSM,keller2004robust, goel2008statistical, amelifard2008multi, knoth2012current, Shahin-TVLSI2011,  fatemi2006statistical, fatemi2007current} 
use voltage-dependent components to model logic cells. 
In addition to higher accuracy, another advantage of current based models over voltage based models is the ability to simulate realistic output waveforms for arbitrary input signals. 
The major shortcoming of LUT-based approaches is the high latency for memory queries.

In this work, we present NN-PARS, a \textit{neural network} (\textbf{NN}) based \textbf{PAR}allelized circuit \textbf{S}imulation framework that replaces current based CSM LUT queries with NN computations and exploits the architecture of \textit{graphical processing units} (GPUs) for concurrent simulation. 
By following our proposed method, various gates in the circuit can be simulated in parallel. 
An event-driven scheduling engine is embedded that selects gates for computation based on characteristics of the underlying GPU platform and the input netlist to minimize the total circuit simulation time. 
The major novelties of our NN-PARS framework are as follows:
\begin{itemize}
    \item NN-PARS accelerates the CSM simulation of complex integrated circuits using optimized NN structures considering the underlying GPU computational capabilities.
    \item Considering the iterative nature of output signal waveform calculation based on CSM, NN-PARS embeds a simple event-driven scheduling methodology to further maximize simulation concurrency by performing calculation steps for many logic cells in the circuit in parallel, hence disentangling logic cell simulation from the order of cells in the circuit topology.  
\end{itemize}
The remainder of our paper is organized as follows. Section~\ref{sec:background} presents a brief background on CSM simulation. Sections~\ref{sec:methodology} and ~\ref{sec:experiments} elaborate our NN-PARS framework and experimental results, respectively. Section~\ref{sec:conclusion} concludes the paper. 
\vskip 5mm

%% file: background.tex
\section{Background} \label{sec:background}

Although our NN-PARS framework can be utilized to enhance any LUT based circuit simulation technique, we choose CSM as the method of comparison.
CSM technique models each logic cell with voltage-dependent current sources, as well as input, miller, and output capacitors~\cite{croix2003blade, fatemi2007current, fatemi2006statistical}. 
In the case of a simple INV gate, CSM components are only dependent on input ($V_{I}$) and output ($V_{O}$) voltages.
However, for logic cells with multiple numbers of inputs, these components depend on a larger number of variables, i.e voltage of inputs and internal nodes~\cite{amelifard2008multi}. 
Consequently, the size of CSM LUTs grow exponentially with the number of variables.

\begin{figure}
\centering
\includegraphics[width = 0.48\textwidth, keepaspectratio] {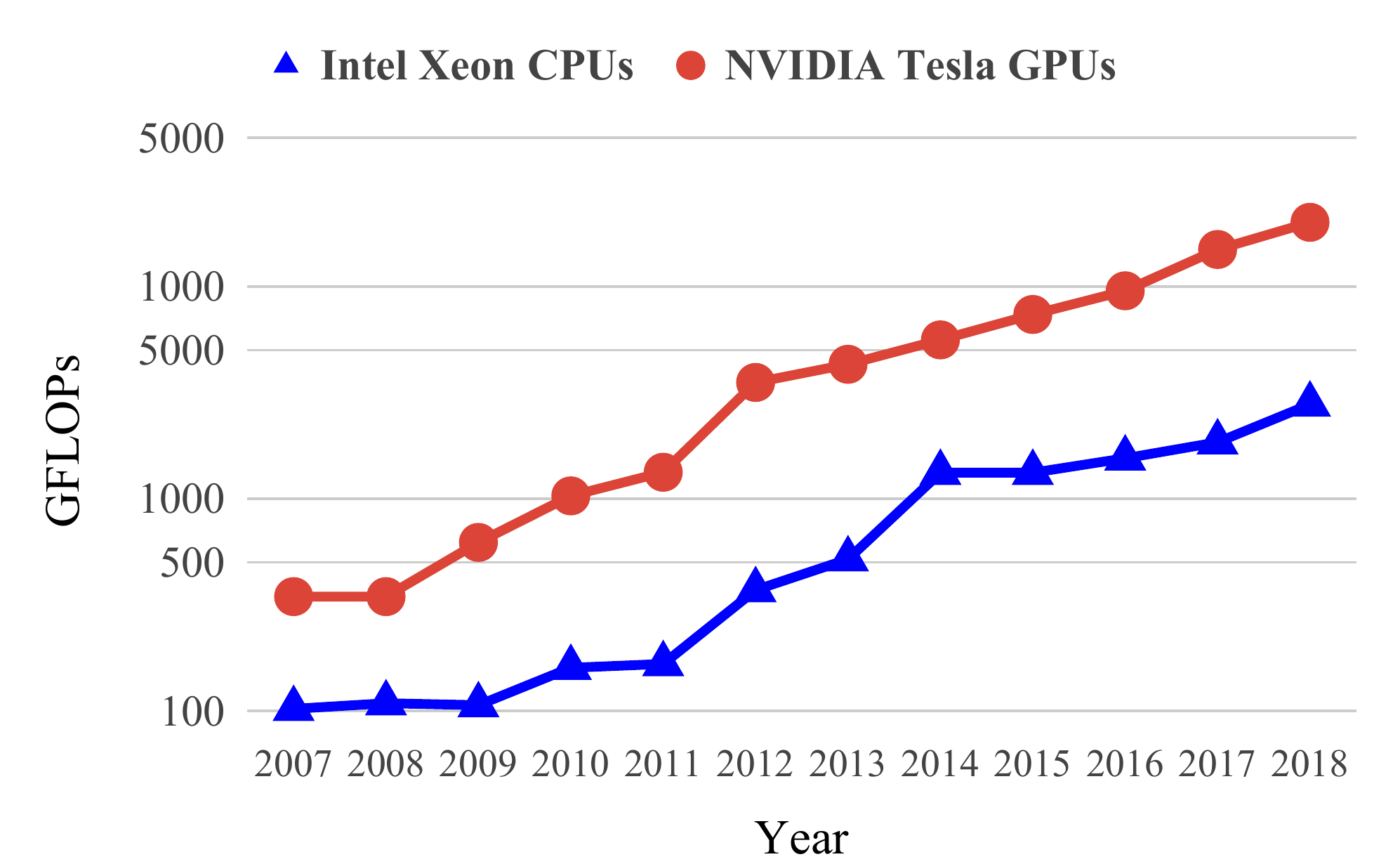}
\caption{The evolution of computational capabilities of CPUs and GPUs in terms of peak single precision floating point operations per second (FLOPS). Vertical axis is depicted in logarithm scale. }
\label{fig:GPU-trend}
\end{figure}

Despite the recent advances in computational capabilities of CPUs, such as process parallelization by introducing many-core processors with dedicated cache memory, they still lack high efficiency when processing tasks with a large number of parallel computational sub-tasks. 
GPUs are specifically designed to outperform CPUs for such tasks with capability of higher order parallel computation. 
Moreover, these devices are also known as an efficient hardware platform for training and inference of NN~\cite{Ng-ML-GPU}.
This is partly because of two levels of parallelized processing units in GPUs: several \textit{multiprocessors} (MPs), and several \textit{stream processors} (SPs, also referred as cores) that run the actual computation for each multiprocessor. 
Each core is equipped with arithmetic units, register files, and designated cache. 
The superiority of GPUs can be observed by comparing the evolution of GPUs and CPUs in terms of \textit{number of floating-point operations per second} (FLOPS) as shown in Fig.~\ref{fig:GPU-trend}. 

As suggested in~\cite{csm-iccd}, high dimensional CSM-LUTs with large sizes can only fit in DRAM of CPUs or GPUs, while low-dimensional V-LUT tables can easily fit into L1 caches. 
The major shortcoming in data retrieval from DRAM is the high latency. 
As an example, specification of a 24-core Intel processor with Broadwell microarchitecture~\cite{Broadwell} given in Table~\ref{tab:cpu-spec} shows that the memory access in DRAM is about 2 orders of magnitude slower than that of L1 cache. 
Another disadvantage of memory query is that in contrast to the dedicated caches for each core in multi-core processors and GPUs, the main memory is shared. 
However, the number of parallel reads from DRAM to processors, referred to as memory-channel, is limited and is much lower than the number of cores. As an example, the 24-core processor in Table~\ref{tab:cpu-spec} has only 4 memory channels. 


Dependency on memory drastically increases the total circuit simulation time and specially prevents accurate approaches such as CSM to be practical. 
To mitigate this shortcoming, semi-analytical methods~\cite{csm-cui-aspdac2014} suggest combining nonlinear analytical models and low-dimensional CSM lookup tables to simultaneously achieve high modeling accuracy in addition to low time and space complexity. 
On the other hand, ~\cite{csm-iccd} (referred to as CSM-NN method throughout the paper) proposed complete removal of the long memory queries by approximating CSM component values using simple NNs. 
While this method improved the simulation time of simple gates, it did not touch upon on how it can be scaled up to the level of circuit simulation, especially using parallel computation capabilities of GPUs. 

In the following two sections, we present how our NN-PARS parallelizes simulation of logic cells in a circuit, while avoiding high latency memory retrievals needed in LUT based simulators, and further speeds up the simulation process by scheduling the concurrent tasks according to the GPU processing capabilities.

\begin{table}
\centering
\caption {Latency values for information retrieval from different hierarchy levels of memory and hardware specifications of Intel Xeon E7-8894 v4 server processor with Intel Broadwell microarchitecture. The computational capability of the processor is given in \textit{Giga floating point operations per seconds} (GFLOPs).} \label{tab:cpu-spec} 
\begin{tabular}{lll} \hline \hline
\multicolumn{3}{c}{\textbf{Intel Broadwell Microarchitecture}} \\ \hline 
\textbf{Memory} & \textbf{Size (KByte)}     & \textbf{Latency (Clock Cycle)} \\ 
L1 Data Cache   & 32 per core     &  4-5\\
L2 Cache        & 256    &   11-12 \\
L3 Cache        & 60,000  & 38-42 \\
DRAM            & -  & $\approx$ 250  \\ \hline \hline
\multicolumn{3}{c}{\textbf{Intel Xeon Processor E7-8894 v4}} \\ \hline 
\multicolumn{2}{l}{Cores} & 24 \\
\multicolumn{2}{l}{Base Frequency} & 2.40 GHz \\
\multicolumn{2}{l}{Theoretical Peak Computation}  & 920 GFLOPs \\
\\
\end{tabular}
\end{table}

%% file: method.tex
\section{NN-PARS Framework}\label{sec:methodology}
The characterization in this method is the same as conventional CSM-LUT. 
We followed the same training flow, i.e. choice of network architecture, optimization algorithm, preprocessing, and evaluation as in CSM-NN.
The following section explains modeling the CSM of standard cells with NN, required resources for parallel computation of NNs and latency on GPU platform, and finally the flow of circuit simulation, including the event-driven scheduling of NN-PARS.

\subsection{NN Architecture} \label{sec:neural-network}
We followed the same approach as in CSM-NN to substitute memory retrieval with NN computation for simple logic cells. 
Every logic cell in the library is modeled by a NN with one single hidden layer.

It is very important to note that while accuracy of NNs in predicting CSM component values is important, the accuracy should ultimately be reported based on the quality of the output waveforms, and not just a certain measurement such as logic cell delay. This coincides with the functionality of CSM in regenerating circuit voltage waveforms. 
Therefore, similar to~\cite{csm-iccd} and~\cite{ssta-csm-dac2016}, we use the \textit{expected waveform similarity} ($\mathrm{E_{sim}}$) as a figure of merit for the simulation accuracy measurements. 
In this work, $E_{sim}$ is defined as the mean of the absolute difference between precise HSPICE and NN-PARS simulations relative to the supply voltage value of the technology as shown in Eq.~\ref{eq:esim}. 

\begin{align}\label{eq:esim}
    \mathrm{E_{sim} = \frac{1}{T \times V_{DD}} \int_{0}^{T} |V_{ SPICE}-V_{ NN-PARS}|}
\end{align}

In addition to a measurement for reporting the accuracy of the results, we used $E_{sim}$ to find the architecture of NNs. 
The smallest number of neurons such that the model can pass a pre-defined accuracy threshold in terms of $E_{sim}$ when stimulated with set of noisy inputs is selected for the NN implementation of the logic cell. 

\subsection{Computational resources and latency analysis}
The main advantage of this proposed method is the high parallelizability and consequently very low latency in simulation of circuits when computed on GPU platforms. 
Therefore, a detailed analysis of the latency and the number of required computation resources of the CSM-NN is necessary. 
The main computational operations of a single-hidden-layer NN are multiplication (MUL) and addition (ADD). 
GPU cores are designed to perform one MUL and one ADD in a single cycle~\cite{cuda-core}. 
Considering the number of inputs and size of the hidden layer as $D$ and $H$ respectively, there are $D \times H$ multiplications in the first layer. 
It is very important to note that there are no dependencies among MUL operations in one specific layer, therefore they can all be computed in parallel using $D\times H$ cores within a single cycle. 
We occupied these \textit{initial cores} in this cycle, but they can be reused in the next cycles. 
To calculate the output of each of the $H$ hidden neurons, $D$ values should be accumulated to generate the output. 
This can be efficiently parallelized by using tree-structures within $\ceil{log_2(D+1)}$ cycles. 
The number of required cores in the first cycle is $\frac{D+1}{2}$, which is less than the number of initial cores, thus no further core allocation is required and the computation can be done on initial ones. 
Following the same approach for the output layer, we can conclude that single-hidden-layer NN can be computed with $D \times H$ cores within latency given in Eq.~\ref{eq:latency}. 

\begin{align}\label{eq:latency}
    \text{Latency } &= 1 + \ceil{log_2(D+1)} + 1 + {\ceil{log_2 (H+1)}}
\end{align}

By implementing a trained NN with fixed parameters on a GPU, the weights of each operation can be stored in register files, therefore, there is no need to retrieve data from memory.

\begin{table}
\caption {NVIDIA TESLA V100 GPU Specifications.} \label{tab:gpu-spec} 
\centering
\begin{tabular}{ll}
\hline
Streaming Processors (SM)           & 80 \\
32bit FP CUDA core (per SM/total)   & 64/5120 \\
64bit FP CUDA core (per SM/total)  & 32/2560 \\
Register files per SM & 256/4~KB \\
L1 cache / shared memory (per SM/core) & 128/2~KB \\
L1 cache hit latency: & 28 \\
Base clock frequency   & 1450~MHz \\
Single precision FLOPS & 14.8~TFLOPS \\ \hline
\end{tabular}
\end{table}

\subsection{Concurrent simulation of gates in CSM}
In CSM simulation, voltage waveform calculation is performed in a series of short time intervals ($dT$) in an iterative process. 
Considering the voltage values ($V_I$s) and input slews ($\Delta V_I$) are known for all gates in one interval ($T_i$), the change in voltage can be calculated for the next interval ($T_{i+1}=T_i+dT$). 
In other words, the change in output voltage of a driver gate $DG$ in one time interval $(\Delta V_{O}^{(D)}(t_{i}))$, is the input voltage change of the load gate ($L$) in the next time interval $(\Delta V_{I}^{L}(t_{i+1}))$.
Following this approach, the simulation of gates in a single interval are not dependent to each other and can potentially be done in parallel.
On the other hand, voltage based simulation calculates the delay and output slew of a single gate based on the input slew, i.e. output slew of the driver gate, and the capacitive load.
The dependency of delay calculation of load gates to simulation of their driver gates, prevents voltage based methods to simulate gates from different \textit{levels} of the circuit in parallel. 

\subsection{NN-PARS circuit simulation flow}
\begin{figure}
    \centering
    \includegraphics[width=0.6\linewidth]{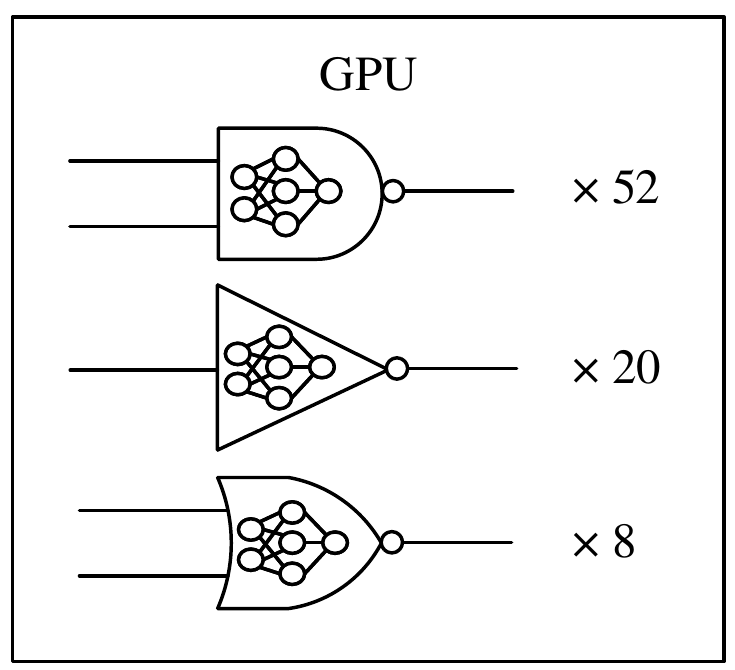}
    \caption{Number of NN models implemented on Tesla V100 GPU (Table~\ref{tab:gpu-spec}) from each cell in the library (INV, NAND2 and NOR2) for optimal NN-PARS simulation of C7552 circuit in FinFET-7nm technology.}
    \label{fig:gpu-pars}
\end{figure}

To better illustrate the steps of our NN-PARS, we use C7552 netlist from ISCAS85~\cite{ISCAS85} benchmark as an example circuit and the GPU platform introduced in Table~\ref{tab:gpu-spec} as an example processor. 
To further simplify our description, we limit the standard cells to INV, NAND2, and NOR2. 
First, NN-PARS identifies the count of gates from each standard cell in the circuit netlist. For example, there are 2625 NAND2, 799 INV, and 401 NOR2 gates in C7552. Based on the relative ratio of these counts, we dedicate GPU cores to model the cells as shown in Fig.~\ref{fig:gpu-pars}. 
Now that all the computational cores of GPU are dedicated, we can start the circuit simulation.
A simple event driven simulation scheduler is designed that schedules the steps of simulation.
According to the number of models on GPU for each cell, random gates in the circuit are selected for simulation. 
Due to independency at each interval, CSM simulation can be performed in parallel for many gates. Thus, at each time interval, NN-PARS selects a subset of gates to run on GPU and simulate. 

In our example, at each time frame, 52, 20 and 8 NAND2, INV and NOR2 gates of the circuit can be simulated in parallel (c.f. Fig.~\ref{fig:gpu-pars}). Similar to this subset, all other gates are simulated for this time interval. This means that for C7552, it takes GPU $3825/80$ iterations to  simulate the circuit for one time interval. 

Although CSM simulation of a logic cell at a certain time interval does not depend on that of other logic cells in that time interval, random selection of logic cells as a subset to be simulated on the GPU may not be optimal. This is because, in fairly large circuits, a large number of cells do not require any simulation in one time interval as their voltage levels for different nodes were not changed in the previous one.  
Therefore, the event driven simulation scheduler of NN-PARS  neglects the unnecessary gate simulations. 
The NN-PARS scheduler assigns the logic cells with voltage values changed beyond a threshold to the active set so they will be simulated in the next time frame. 
On the other hand, the logic cells with no changes in any of their voltage nodes are removed from the active set. 

%% file: experiment.tex
\section{Experiments and Simulation Results} \label{sec:experiments}

We implemented the simulator and the flow of our NN-PARS framework in Python. 
Our implementation is technology independent and can characterize, and create NN models with flexible configurable setups, for various logic cells. 
More importantly, the simulator can exploit GPU in order to parallelize the simulation of the given combinational circuit netlist.
NN implementation and training are based on the Scikit-learn~\cite{scikit-learn} package. 

CPU and GPU devices that are used as platforms for CSM and NN-PARS are introduced in Table~\ref{tab:cpu-spec} and Table~\ref{tab:gpu-spec} respectively. 
Hardware platforms are comparable to each other in terms of cost (about 8,000 USD) and the production year (2017) in order to have a fair comparison. 

\subsection{Selected Technologies}
For better evaluation of our NN-PARS and its technology independence characteristics, we performed our experiments on both MOSFET (16nm) and FinFET (7nm) devices from Predictive Technology Model~(PTM)~\cite{PTM} packages. 
Two device types namely \textit{low-standby power} (LP) and \textit{high performance} (HP) are used in our experiments~\cite{finfet-msa}.

\subsection{Training for logic cells}
The total number of generated data points by characterization is 500 samples per gate.
The data was randomly split into training (90\%) and test (10\%) datasets. 
The exponential range of the $I_O$ values (from $pA$ to $\mu A$) is not optimal for training nonlinear regression models. 
Therefore, we trained our models on $log I_O$ values. 
The normalization of data in regression problems would help the solvers with faster convergence and better numerical stability. 
This process is implemented inside our solver~\cite{scikit-learn}. 
To select the optimal size of the hidden layer for each model, we repeated the training process for various neuron numbers in the range of $5-40$. 
Each of the trained models was tested by applying a set of noisy input signals. 
The model with the minimum size of the hidden layer that met $E_{sim} < 1 \%$ threshold is chosen as the NN-PARS architecture for the logic cell. 
The complete results for the choice of architecture for INV, NAND2, and NOR2 NN-PARS models are given in Table~\ref{tab:cell-NN}. 

\begin{table}
\centering
\caption {Choice of NN hidden layer size for single and two input logic cells.} 
\label{tab:cell-NN}
\begin{tabular}{|l|c|c|c|} \hline
               & INV  & NAND2 & NOR2  \\  \hline
MOSFET-HP 16nm & 9    & 18 & 18  \\  \hline
MOSFET-LP 16nm & 8   &  17 &  17 \\  \hline
FinFET-HP 7nm  & 10   &  20   & 20 \\  \hline
FinFET-LP 7nm  & 10   &  21  & 21 \\ \hline
\end{tabular}
\end{table}

\subsection{Circuit Simulation}
\begin{figure}
\centering
\includegraphics[width=0.48\textwidth, keepaspectratio] {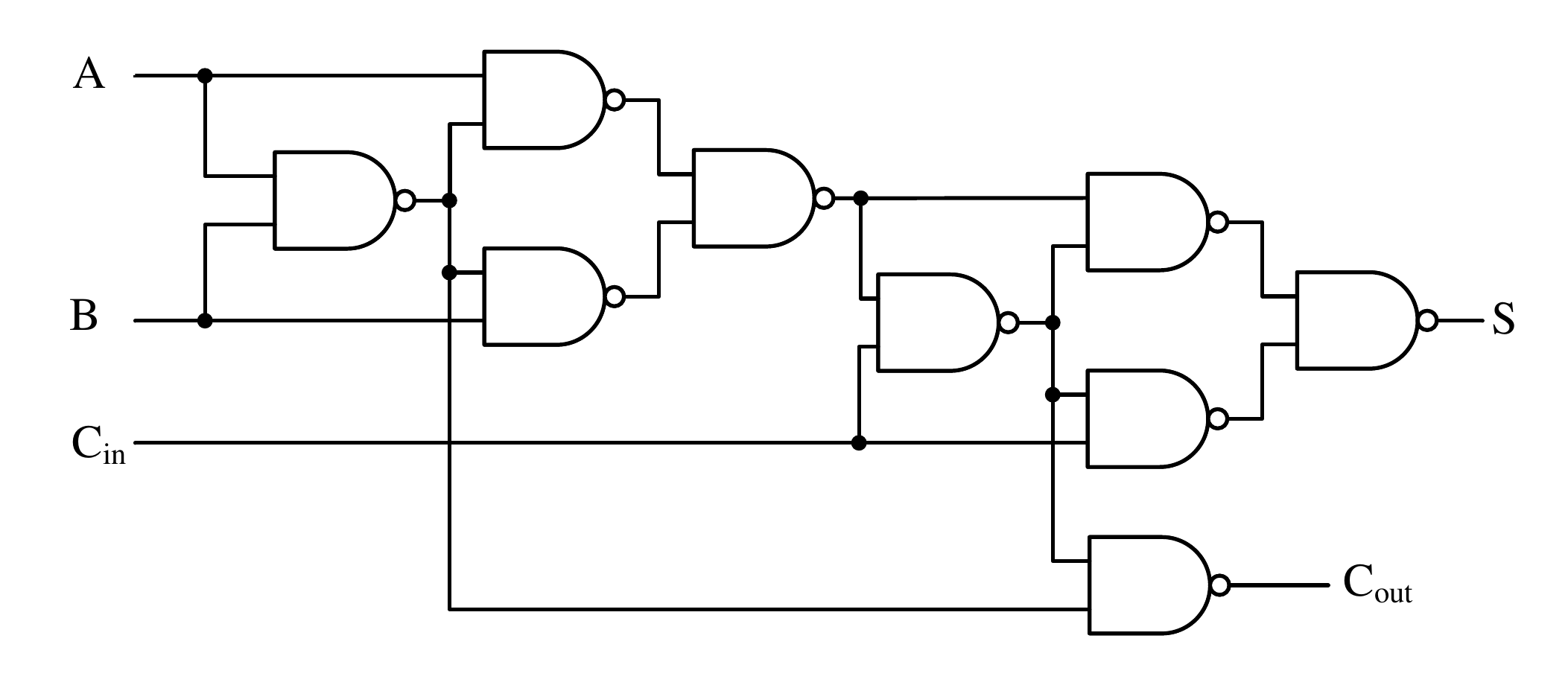}
\caption{Gate level schematic of the full adder circuit used in our experiments.}
\label{fig:full-adder}
\end{figure}

In this work we evaluated our NN-PARS framework by simulating a \textit{full adder} (FA) circuit with schematic shown in Fig.~\ref{fig:full-adder}. In addition, we analysed the performance improvement achieved by NN-PARS compared to NN-LUT for real combinational circuits from ISCAS85 benchmarks~\cite{ISCAS85}. 

\begin{figure}
\centering
\includegraphics[width=0.48\textwidth, keepaspectratio]{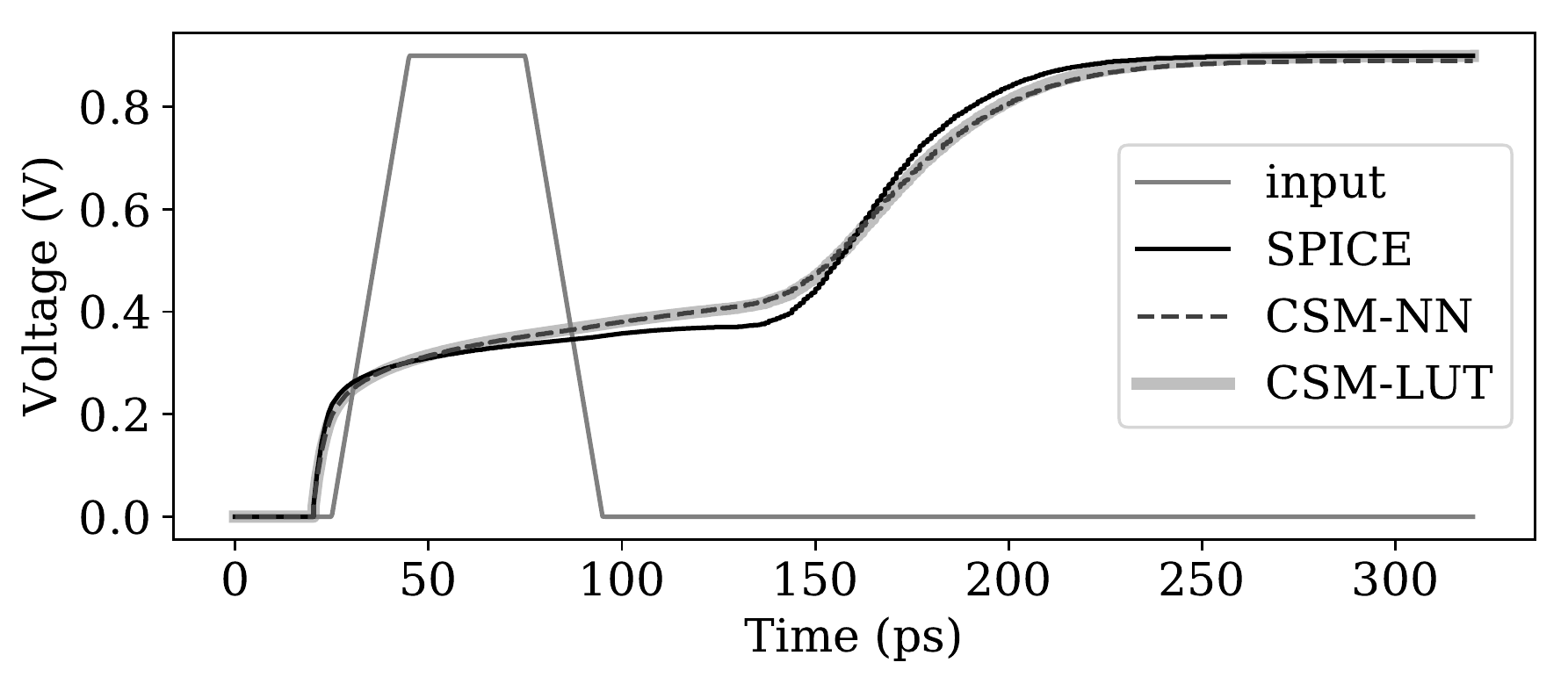}
\caption{Waveform of full adder simulation with SPICE, CSM-LUT and NN-PARS.}
\label{fig:waveform}
\end{figure}

CSM-LUT method is considered to be computed on the CPU platform as it does not benefit from GPU parallelization. 
The required computation resources and latencies for GPU implementation of NN-PARS are calculated using equations in Section~\ref{sec:neural-network}. 
Comparing the output waveforms of SPICE, CSM-LUT, and NN-PARS methods in Fig.~\ref{fig:waveform} confirm the simulation accuracy of NN-PARS. 
We also measured $E_{sim}$ by comparing output waveforms of HSPICE as the baseline with those of NN-PARS simulations. Results in Table~\ref{tab:res-FA} suggest that $E_{sim}$ is limited to 2\%. 

\begin{table}
\centering
\caption{CSM simulation results of a full adder circuit in both FinFET and MOSFET technologies. The simulation time improvements is the ratio of the time required for CSM simulation over the one for NN-PARS. The hardware platform specs are reported in Table~\ref{tab:cpu-spec} and Table~\ref{tab:gpu-spec}. $E_{sim}$ is the measure of accuracy introduced in Eq.~\ref{eq:esim}. }
\label{tab:res-FA}
\begin{tabular}{|c|c|c|c|c|} \hline
Technology & \multicolumn{2}{c|}{MOSFET 16nm} & \multicolumn{2}{c|}{FinFET 7nm} \\ \hline 
Device & HP & LP & HP & LP \\ \hline
$E_{sim}$   & 1.64\%  & 1.27\% & 1.81\%  & 1.77\% \\ \hline
\textit{Improvement} & \multicolumn{4}{c|}{30.4} \\ \hline
\end{tabular}
\end{table}

As we can see in Table~\ref{tab:res-FA}, the improvement achieved by NN-PARS is the same for different devices as all the gates of the FA can be modeled on our GPU in parallel. 
The limited number of gates in the FA circuit does not reveal the full performance increase of NN-PARS. 
Therefore bigger circuits with thousands of gates were analyzed. The results are reported in Table~\ref{tab:iscas85-res}. 

\begin{table}[]
\centering
\caption{Improvement of simulation time in NN-PARS for combinational circuits from ISCAS85 benchmarks~\cite{ISCAS85}}
\label{tab:iscas85-res}
\begin{tabular}{|c|c|c|c|}
\hline
-     & \# gates & MOSFET  & FinFET \\ \hline
c880  & 383     & 92$\times$        & 81$\times$       \\ \hline
c1355 & 546     & 120$\times$       & 124$\times$       \\ \hline
c7552 & 3825    & 134$\times$       & 134$\times$       \\ \hline
\end{tabular}
\end{table}

%% file: conclusion.tex
\section{Conclusions}\label{sec:conclusion}

Our goal in this work was to resolve the accuracy and latency issues of existing simulation methodologies that heavily depend on memory queries. Our NN-PARS framework replaces long memory queries with efficient and parallelizable NN based computations and employs an optimized event-driven scheduling engine that concurrently runs the simulation events of logic cells in the circuits. 

The simulation latency of NN-PARS was evaluated in multiple MOSFET and FinFET technologies based on predictive technology models. The results confirm that NN-PARS improves the simulation speed by up to $134\times$ compared to a state-of-the-art current based CSM baseline in large circuits. Furthermore the high accuracy of NN-PARS in terms of waveform similarity was evaluated w.r.t. HSPICE. We expect the application of NN-PARS in analysis and optimization of advanced VLSI circuits such as system-on-chips (SoCs) will significantly improve the quality of results.
\vspace{1mm}